\documentclass[twocolumn,showpacs,preprintnumbers,amsmath,amssymb,superscriptaddress,pra]{revtex4-1}

\usepackage{graphicx}
\usepackage{amsmath}
\usepackage{amssymb}
\usepackage{bm}

\newcommand{\na}{$^{23}$Na~}

\begin{document}
\title{Fast production of ultracold sodium gases using light--induced desorption and optical trapping}
\author{Emmanuel Mimoun}\email{emmanuel.mimoun@ens.fr}
\affiliation{Laboratoire Kastler Brossel, ENS, UPMC, CNRS, 24 rue Lhomond, 75005 Paris}
\author{Luigi De Sarlo}
\affiliation{Laboratoire Kastler Brossel, ENS, UPMC, CNRS, 24 rue Lhomond, 75005 Paris}
\author{David Jacob}
\affiliation{Laboratoire Kastler Brossel, ENS, UPMC, CNRS, 24 rue Lhomond, 75005 Paris}
\author{Jean Dalibard}
\affiliation{Laboratoire Kastler Brossel, ENS, UPMC, CNRS, 24 rue Lhomond, 75005 Paris}
\author{Fabrice Gerbier}\email{fabrice.gerbier@lkb.ens.fr}
\affiliation{Laboratoire Kastler Brossel, ENS, UPMC, CNRS, 24 rue Lhomond, 75005 Paris}
\date{\today}
\begin{abstract}
 In this paper, we report on the production of a Bose--Einstein condensate (BEC) of sodium using light--induced desorption as an atomic source. We load about $2\times 10^7$~atoms in a MOT from this source with a $\sim6$~s loading time constant. The MOT lifetime can be kept around $27$~s by turning off the desorbing light after loading. We show that the pressure drops down by a factor of 40 in less than $100$~ms after the extinction of the desorbing light, restoring the low background pressure for evaporation. Using this technique, a \na BEC with $10^4$ atoms is produced after a $6$~s evaporation in an optical dipole trap.
\end{abstract}
 \maketitle
%
%

\section{Introduction}

Sodium is one of the first atomic species which has been brought to quantum degeneracy~\cite{davis1995a}. Among the alkalis, it allows to produce the largest Bose--Einstein condensates (BECs) with atom numbers $>10^8$~\cite{streed2006a,stam2007a}. This is in part due to the efficiency of laser cooling, but also to favorable collisional properties (large elastic cross--section, low inelastic losses). These properties allow efficient evaporative cooling to Bose--Einstein condensation~\cite{streed2006a}, as well as the production of large degenerate Fermi gases ({\it e.g.} $^6$Li) when the sodium cloud is used as a buffer gas for sympathetic cooling~\cite{hadzibabic2003a}.  Sodium also has antiferromagnetic spin--dependent interactions~\cite{ohmi1998a,KetterleNobel}, which can lead to complex entangled spin states~\cite{law1998a}. Such spin states are particularly sensitive to stray magnetic fields~\cite{Ho2000a,Castin2001,Ashhab2002a}, so that they are expected to survive only in a quiet magnetic environment provided by magnetic shielding. Hence their study, which is the main motivation behind this work, requires to optimize the production of cold sodium gases in a compact setup compatible with such shielding. 

A typical ultracold gas experiment can be decomposed into three successive steps : a source delivering hot atoms to the vacuum chamber, a magneto--optical trap (MOT) capturing atoms from the source and pre-cooling them, and a conservative trap loaded from the MOT where evaporative cooling is performed to reach quantum degeneracy. The MOT is common to all experimental setups, which differ in the first and last steps. The atom source can be an atomic beam produced by a Zeeman slower~\cite{phillips1982a,Barrett1991a}, or a residual vapor that exists in the ultra--high vacuum (UHV) chamber due to, {\it e.g.} a nearby atomic reservoir~\cite{monroe1990a,prentiss1988a} or atomic dispensers~\cite{fortagh1998a,ott2001a,hansel2001a}. The conservative trap can be either a magnetic trap for spin--polarized atoms, or an optical dipole trap for unpolarized mixtures. All practical implementations must solve an intrinsic quandary: on the one hand, the atom flux in the MOT region must be large enough for efficient loading, and on the other hand, it has to be low enough that collisions with the background vapor do not hinder evaporative cooling in the conservative trap. This has been solved in various ways, by spatially separating the MOT capture region and the evaporative cooling region, or by modulating the atomic density in time (the Zeeman slower technique, combined with a controllable beam block, is an example of the latter solution). 

\begin{figure*}[t]	
\centering{\includegraphics{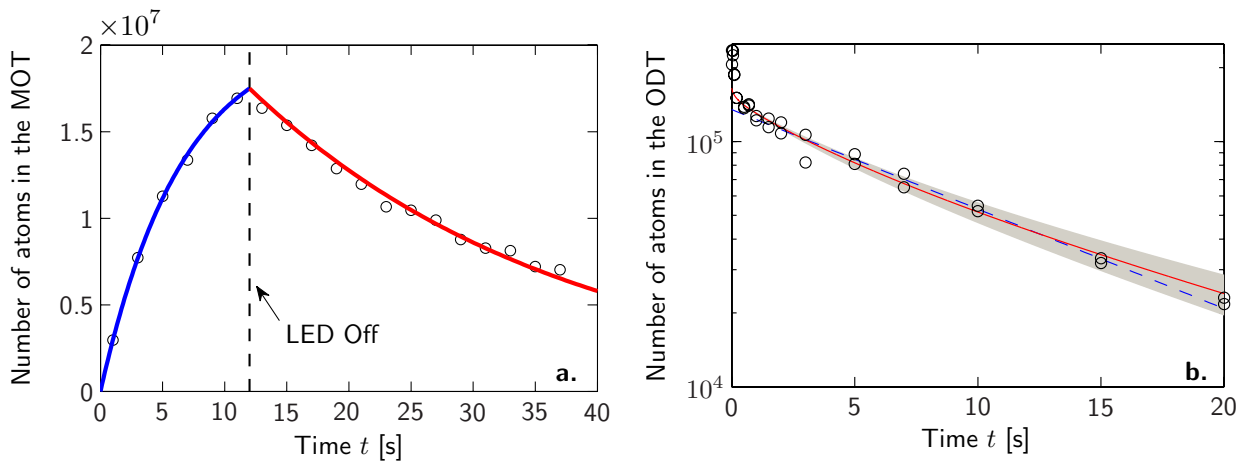}}
\caption{(Color online) ({\bf a}): MOT loading and decay dynamics. The MOT is loaded using LIAD at maximum power for $12$~s. The desorbing light is then turned off. We find $1/e$ time constants of $\tau_{\rm ON}=6.5\pm 0.5$~s and $\tau_{\rm OFF}=27\pm 1.5$~s for the loading and decay, respectively, and an asymptotic number of atoms $N_{\rm st}=2.1\times10^7$~atoms. ({\bf b}): Decay of the number of atoms held in the optical dipole trap (ODT). A simple exponential fit yields a decay time $\tau_{\rm ODT}=10.7\pm 1$~s (dashed blue line) for the data corresponding to $t>200$~ms. A model including 3--body losses with the rate measured in~\cite{gorlitz2003a} has been used to fit a one--body collision rate $\tau_{\rm 1B}=17\pm 3$~s for the same data (solid red line, 95\% confidence bound in gray), see Sec.\ref{DTLt} for details.} 
\label{fig1} 
\end{figure*}

In order to reduce the complexity and size of the apparatus, the second solution seems more favorable, provided that one is able to truly switch on and off the vapor pressure in the UHV chamber. Atomic dispensers, which allow to release an alkali vapor by thermally activating the reduction of an inert alkali metal oxide, have been introduced for this purpose in connection with atom chips experiments~\cite{fortagh1998a,ott2001a,hansel2001a}. Unfortunately, dispensers are unsuitable to modulate the pressure inside the chamber with time constants below one second due to their thermal cycle~\cite{fortagh1998a}. This forces one to work at a ``compromise'' pressure, which allows one to reach BEC thanks to the high collision rates obtained in atomic chips~\cite{ott2001a,hansel2001a}, but limits the sample lifetime. 

A promising technique to rapidly modulate the atomic pressure is light--induced atomic desorption (LIAD)~\cite{Gozzini1993a,Anderson2001}, a phenomenon analogous to the photoelectric effect in which an adsorbed atom is released from an illuminated surface by absorbing a photon. LIAD was initially observed in coated \na cells and subsequently studied for different atomic species and several substrates~\cite{Gozzini1993a,meucci1994a,Xu1996Na,Atutov1999,Alexandrov2002,Karaulanov2009,Bogi2009}. Its usefulness as a source for quantum gases experiments was demonstrated for rubidium~\cite{Anderson2001, hansel2001a, Atutov2003, Du2004, aubin2005a} and potassium~\cite{Aubin2006a, Klempt2006}.

To our knowledge, all the experimental groups which have been able so far to produce a \na\ BEC relied on a Zeeman slower to load the MOT~\cite{davis1995a,hau1998a,naik2005a,dumke2006a,stam2007a}. Furthermore, with the exception of Ref.~\cite{dumke2006a}, where evaporative cooling is done in an optical trap, these experiments used magnetic trapping. When considering only the MOT loading step, which is less restrictive than evaporative cooling in terms of acceptable pressure, several alternatives have been demonstrated~\cite{prentiss1988a,cable1990a,milori1997a}, but are difficult to operate under UHV conditions. Loading from \na dispensers was also demonstrated~\cite{muhammad2008a}, but not under conditions suitable for achieving BEC. Only very recently a group has reported the use of LIAD for loading a \na MOT~\cite{telles2009a}. In that work, the atomic pressure drops after the extinction of the desorbing light to a level (lifetime $\sim 8~$s) that could be compatible with evaporative cooling. 

Here we report the experimental realization of an ultracold \na gas in a single UHV chamber, where LIAD is used to increase the \na pressure for MOT loading and where the atoms are captured in an optical dipole trap for evaporative cooling. We achieve (see Fig.(\ref{fig1}{\bf a})) a loading of the MOT with a time constant $\sim 6~$s with LIAD on, a subsequent lifetime of the MOT of $\sim 27$~s with LIAD off, and a lifetime in the dipole trap with a time constant $\sim11~$s. The latter is partially limited by evaporation and inelastic collisions between trapped atoms. In our setup, the ratio between the partial pressure of sodium when LIAD is switched on and off is $\eta\simeq 40$. The delay for observing this pressure drop after the switching off of LIAD is less than $100$~ms. This allows us to quickly switch from MOT loading to evaporative cooling. Under these conditions, we observe BECs containing $\sim 10^4$ atoms after a $6$~s evaporation. The paper is organized as follows. In Section~\ref{sec:experiment}, we describe our experimental setup. We present the experimental results on MOT loading using LIAD in Section~\ref{sec:LIAD}. We compare our results to other experiments in Section~\ref{sec:disc}. 

\section{Experimental setup}\label{sec:experiment}
\subsection{Vacuum system}

\begin{figure*}[t]
\centering{\includegraphics{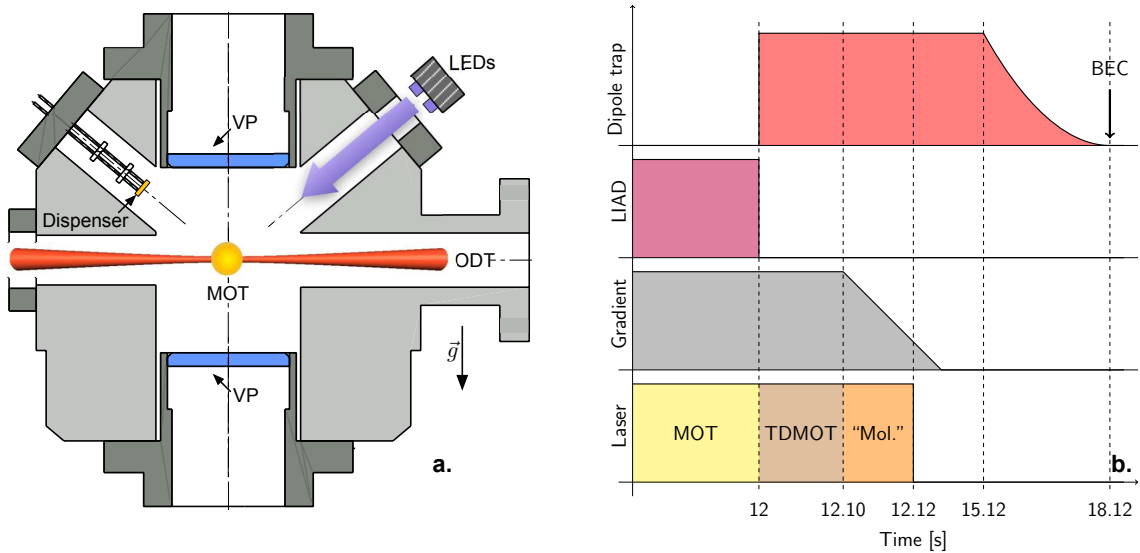}}
\caption{(Color online) ({\bf a}): Sketch of the UHV chamber, showing the location of one of the pairs of LEDs with respect to the MOT and the reentrant viewports (VP), as well as the path of one of the ODT arms. ({\bf b}): Time sequence of the experiment. We show from the top down the power of the ODT, the current in the LEDs, the magnetic gradient of MOT coils and the different cooling phases. TDMOT stands for Temporal Dark MOT and ``Mol.'' for ``optical molasses'' (see Sec.\ref{ODT}).}
\label{fig2}
\end{figure*}

The UHV system is built around a custom--made chamber equipped with several viewports allowing wide optical access. The chamber is made from titanium (for reasons discussed below) machined to a surface roughness specified lower than $700~$nm by the manufacturer (UK AEA Special Techniques, Oxfordshire, UK). The viewports are made from fused--silica windows vacuum--brazed to titanium flanges (MPF Products Inc., Gray Court, SC, USA). The chamber inner volume is about $0.3$~liter. The chamber is connected with CF40 tubing to a getter pump located approximately $20~$cm away from its center, and to a 20~l/s ion pump located approximately $50$~cm away. After 10 days baking at $200^\circ$C to establish UHV, the residual pressure is well below the sensitivity of the ion pump current controller (limited to a few $10^{-9}$~mbar). For all experiments reported in this paper, the pressure near the ion pump stays below the detection threshold of the controller. We estimate the effective pumping speed of the ion pump to be $\sim 4$~l/s in the UHV chamber. 

We have two means to increase the partial pressure of sodium: atomic dispensers and light induced atomic desorption. Atomic dispensers are formed by a powder of alkali oxide, which is chemically inert at room temperature
. A gas of alkali atoms is released from such devices by activating a chemical reaction with the heat generated by an electric current (typically a few amperes) running through the metallic envelope containing the powder. The dispensers used in our experiment (Alvatec GmbH, Althofen, Austria) are run at a relatively low current ($<4~$A) to avoid release of large sodium loads into the chamber. A pair of such dispensers is mounted using standard vacuum connectors and a custom CF25 electrical feedthrough (MPF Products Inc., Gray Court, SC, USA) at a distance of about $4$~cm from the center of the chamber. In the first six months after baking out the chamber, the dispensers were flashed about an hour long at $3.6$~A twice a week. We have then found that flashing them for a couple of hours about once every two months was sufficient to maintain a constant number of atoms in the MOT, using LIAD to desorb the atoms from the viewports. Turning on dispensers less often allows to maintain a background pressure at a lower and more constant level.

As an illumination source for LIAD, we use light--emitting diodes (LEDs) emitting near $370-390$~nm (models NCSU033A and NCSU034A from Nichia Corporation, Tokyo, Japan). Each LED is supplied in a small surface--mount chip, and emits around 350 mW of light power. We mount them in pairs on appropriate heat sinks. For the reported experiments, we use two such pairs of LEDs placed around the vacuum chamber. We supply these LEDs with a current ranging from $0$ to $1.5$~A (maximum current). We have verified that the optical power is proportional to the current in this range. 
 
A sketch of the relevant section of the chamber interior is shown in Fig.(\ref{fig2}{\bf a}). Two reentrant flanges supporting large CF63 windows are mounted vertically. The atomic dispensers are mounted close to the viewports using CF25 electrical feedthroughs. The UV LEDs used for desorption are mounted in front of two CF25 viewports. In our case the illuminated surface is partly titanium, and partly a few hundred nanometers--thick layer of alternating TiO$_2$ and SiO$_2$, which constitutes the antireflection coating of our viewports (Duane Mallory, manager of MPF products, private communication, 2009). By changing the position of the UV LEDs, we could experimentally verify that the main contribution to desorption comes from the viewports, and not from the Ti surface. In previous experiments, we also successfully used LIAD to load a MOT in a glass cell (Vycor glass without anti--reflection coating), with similar characteristics as in the present UHV chamber. Attempts made using a 316L stainless steel chamber with small (CF16) viewports failed. In both cases, \na MOTs were also directly loaded from the dispensers.

\subsection{Magneto--optical trap}

We operate the MOT using the all--solid state laser system described in~\cite{mimoun2008a,mimoun2009a}. The laser is locked on the sodium D2 line using modulation transfer spectroscopy on an Iodine cell. We lock on the Iodine P$38$~$(15-2)$ line, which is located $467$~MHz above the sodium D$_2$ resonance frequency~\footnote{This solution has been proposed by Christian Sanner, W. Ketterle's group, MIT.}. The output of the laser is split into two parts. The first part is used to form the main MOT beam and the second to form the repumping beam after passing through a $1.7$~GHz acousto--optical modulator (Brimrose Corporation of America, Sparks, MD, USA). Both beams are delivered to the experiment using single--mode optical fibers. The parameters of the MOT and repumper beams are summarized in Table~\ref{table1}. The MOT is formed using a pair of anti--Helmholtz coils producing a gradient $\sim15~$G/cm on--axis. Three sets of Helmholtz coils are also available to compensate for residual fields during the molasses phase. Typically, we find that only the vertical residual field is significant, with a magnitude compatible with that of the Earth field. We do not observe a significant effect of the magnetic field produced by the dispensers on the MOT when they are fed with current.

\subsection{Optical dipole trap}
\label{ODT}

The optical dipole trap (ODT) is produced using a 20 W fiber laser (IPG Photonics, Oxford, MA, USA). The laser emits in multiple longitudinal modes, and is polarized linearly. We control the beam intensity using a rotating waveplate followed by a Glan--Taylor polarizer, and switch off the laser beam using a fast ($\sim 1$~ms switch--off time) mechanical shutter (Uniblitz, Rochester, NY, USA). This system allows for a stable modulation between $0.5$\% and $100$\% of the laser power within a $60$~Hz bandwidth. The ODT is in a crossed configuration, where the beam is folded onto itself with a 45$^\circ$ angle in the horizontal plane. We took care of making the polarisations of the two arms orthogonal to better than 1$^\circ$. When this was not the case, large heating was observed, presumably due to the fluctuating optical lattices resulting from the interferences between identical frequency components with fluctuating relative phases present in each beam. The useful powers on the atoms are around 14 and 12 W for the first and second arms, respectively~\footnote{This difference is mainly attributed to imperfections of the anti--reflection coatings of the windows, which were apparently damaged during the chamber bake--out.}. We focus the first arm to a $1/e^2$ size of $w_{1}\approx 30~\mu$m and the second arm to $w_{2}\approx22~\mu$m, which corresponds to a depth $V_{0}\approx700~\mu$K for the crossed dipole trap.

For efficient loading of the ODT from the MOT, it is essential to reduce the repumper power by a factor of 50 from its value in the MOT capture phase~\cite{kuppens2000a}. This decreases the steady--state population in the electronic excited state $3$P$_{3/2}$ and increases the spatial density thanks to the reduction of light--induced collisions. We refer to this procedure as ``temporal dark MOT'' (TDMOT) in the following. In our experiment, the ODT is switched on at the beginning of this phase, which lasts for about a hundred milliseconds. It is followed by a 20~ms ``pseudo--optical molasses'' phase similar to the one described in~\cite{dumke2006a}, where the MOT detuning is increased to $36$~MHz and the magnetic gradient is ramped down slowly to zero (see Fig.(\ref{fig2}{\bf b})). During this phase, sub--Doppler cooling helps reducing the temperature. The repumping light is then switched off $1$~ms before the end of the molasses phase, so that all the atoms are optically pumped to the $F=1$ state. We do not perform any Zeeman pumping on the atoms, so that they can be in any state of the $F=1$ Zeeman manifold.

\subsection{All--optical evaporation}
\label{sec:evap}

Starting from about $2\times 10^{5}$~atoms in the crossed ODT, we let the atoms evaporate freely for $3$~s, reaching a phase--space density (PSD) of $\sim 10^{-2}$ (accounting for a factor of 1/3 due to the spin degree of freedom, assuming equipartition in the $F=1$ manifold)~\cite{chang2004a}, with $1.5\times 10^{5}$~atoms at $80$~$\mu$K. We then start ramping down the power of the ODT laser to 1\% of its original value in $3$~s, following a power--law decay as discussed in~\cite{ohara2001a}.  The cloud is then transferred in an auxiliary optical trap which can be turned off fast ($\sim 1$~$\mu$s) using an acousto--optic modulator for time--of--flight imaging. We complete the evaporation ramp to reach Bose--Einstein condensation around $1$~$\mu$K. With further evaporation, we can produce quasi--pure BECs containing $10^{4}$ atoms. The total sequence lasts $18$~s, including $12$~s of MOT loading.

\subsection{Diagnostics}

We infer the properties of the trapped atoms from absorption images. We use two absorption axes, one co--propagating with the first arm of the ODT, the other vertical. We use low--intensity, circularly polarized probe light on resonance with the $F=2\rightarrow F'=3$ transition. A repumping pulse of $\sim300~\mu$s is applied to pump all the atoms in the $F=2$ state before the imaging pulse of $\sim 30\mu$s. The shadow of the atomic cloud is imaged on charge--coupled devices (CCD) cameras, and the inferred density distribution fitted to a Gaussian profile. We measure the number of atoms from the area under the gaussian, using a scattering cross--section $\sigma_{\rm abs}=3\lambda_{0}^2/2\pi$, where $\lambda_{0}\approx589~$nm is the resonant wavelength. We have checked that the absorption by the atomic cloud is peaked around the atomic transition frequency with a $10$~MHz width close to its natural linewidth, and that a linearly polarized probe is absorbed about twice as less, as expected from the square of the Clebsch--Gordan coefficients for the relevant optical transitions. For experiments reported in Sec.\ref{sec:time}, a photodiode monitoring the MOT fluorescence is also used.

\begin{table*}[htbp]
  \centering
  \begin{tabular}{|c|c|c|c|c|c|c|c|c|c|c|c|c|c}
\hline    Beam & Waist & MOT & MOT & TDMOT & TDMOT & Molasses& Molasses\\
     &  &  detuning & power  & detuning & power & detuning & power \\
\hline \hline      
MOT     &  $11$~mm& $-20$~MHz& $1.8$~mW  & $-18$~MHz&$1.8$~mW  & $-36$~MHz&  $2.4$~mW\\
Repumper  & $11$~mm& $0$~MHz& $450~\mu$W  & $0$~MHz &$8~\mu$W  & $0$~MHz&  $8~\mu$W\\
\hline 
\end{tabular}
  \caption{Summary of MOT laser parameters. The waist is the 1/$e^2$ radius of the beam. Laser powers are given per MOT arm and frequency detunings from the $3$S$_{1/2}$ ,$F=2\rightarrow 3$P$_{3/2}$ ,$F'=3$ transition for the MOT beams and the $3$S$_{1/2}$, $F=1\rightarrow 3$P$_{3/2}$, $F'=2$ transition for the repumping beams.}
  \label{table1}
\end{table*}

\section{Magneto--optical trapping of sodium atoms using light--induced desorption}\label{sec:LIAD}
\subsection{Loading the magneto--optical trap}
\label{MOTL}

 We first characterize the loading dynamics of the MOT as a function of illumination. We model the loading dynamics by the equation
\begin{equation}\label{MOTload}
	\dot{N}=R-\frac{N}{\tau_{\rm MOT}}.
\end{equation}                         
Here, $N(t)$ is the number of atoms in the MOT at time $t$, $R$ is the MOT loading rate, and the term $N/\tau_{\rm MOT}$ accounts for losses due to the collisions with the background gas~\footnote{We also tried to model the influence of light--assisted inelastic collisions on MOT loading by adding a term $\beta N^2$ to the loading equation. A fit to the data consistently returned $\beta=0$ for all loading curves we examined, so that we neglect this term in our analysis.}. We assume that $R$ is proportional to the partial pressure of sodium in the chamber $P_{\rm Na}$, while $\tau_{\rm MOT}^{-1}$ is proportional to the sum of $P_{\rm Na}$ and of the residual pressure of each contaminant $i$ present in the vacuum chamber weighted
by the relevant collision cross--sections $\sigma_{{\rm Na}-i}$. We thus write
\begin{equation}
R \approx a P_{\rm Na}, ~~~\frac{1}{\tau_{\rm MOT}} \approx b P_{\rm Na}+\frac{1}{\tau_{0}}.
\end{equation}                         
We take $a$, $b$ and $\tau_{0}$ independent of $N$ and of the illumination, which is a simplifying assumption but describes our data well. Eq.(\ref{MOTload}) then leads to an exponential loading with $1/e$ time constant $\tau_{\rm MOT}$ towards to a steady--state atom number
\begin{equation}\label{eq:Nst}
N_{\rm st} = R\,\tau_{\rm MOT} \approx \frac{a P_{\rm Na} }{b P_{\rm Na}+\frac{1}{\tau_{0}}}.
\end{equation}  
A typical loading is shown in Fig.(\ref{fig1}{\bf a}) (first $12$~s), with $N_{\rm st}=2.1\times10^7$~atoms, $\tau_{\rm MOT}=6.5\pm 0.5$~s and $R=3\times 10^6$~s$^{-1}$. 

\begin{figure}[t]
\centering{\includegraphics{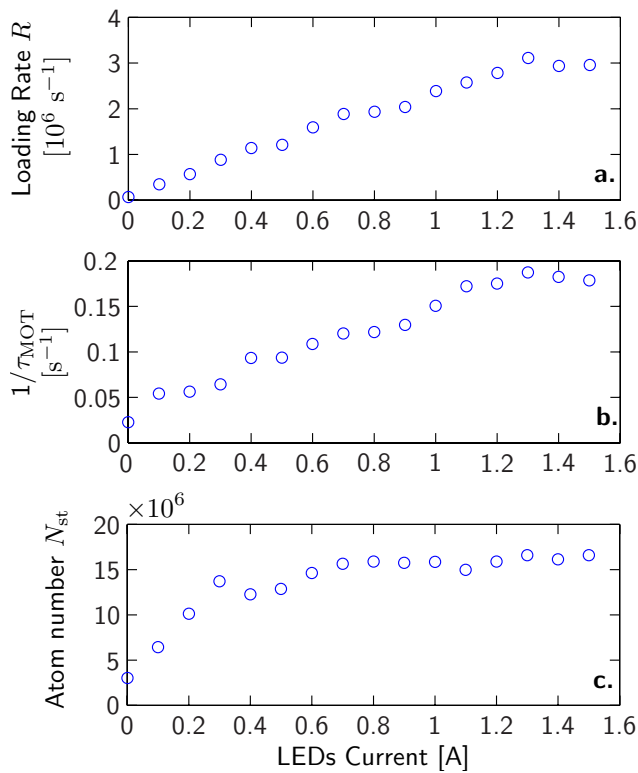}}
\caption{\label{fig3} (Color online) ({\bf a}): Loading rate $R$ of the MOT; ({\bf b}): Inverse of the loading time constant $1/\tau_{\rm MOT}$; ({\bf c}): Steady--state atom number $N_{\rm st}$ for various LED currents. Statistical error bars are smaller than the dots and not shown.}
\end{figure}

A first assessment of the efficiency of the loading of the MOT from LIAD can be achieved by recording the parameters of the MOT as a function of the LEDs current. These measurements are reported in Fig.(\ref{fig3}). The loading rate $R$ (\ref{fig3}{\bf a}) is an increasing function of the currents in the LEDs, showing a growing pressure of sodium in the chamber. We also observe a decrease of the loading times (\ref{fig3}{\bf b}) and an increase of the steady--state atom numbers (\ref{fig3}{\bf c}) when the LEDs current increases.

Eq.(\ref{eq:Nst}) implies that when the loading time is dominated by the sodium partial pressure, the steady--state atom number becomes independent of $P_{\rm Na}$ and therefore of the current in the LEDs. This is what we observe in Fig.(\ref{fig3}{\bf c}) for the highest LED currents. We find $R=3\times 10^6$~s$^{-1}$ when the current in the LEDs is maximal. When the LEDs are off, the loading rate drops to $R=8\times 10^4$~s$^{-1}$. This means that the pressure of sodium with the light switched on is increased by a factor $\eta=40$ with respect the background pressure. 
This factor can be recovered with a relatively good approximation from the value of the MOT lifetime, as we show now. Note first that obtaining an absolute calibration of the pressure is by no means easy in our system, which does not include an UHV gauge. 
When LIAD is off, we expect that $P_{\rm Na}^{\rm off}$ is comparable to the saturated vapor pressure at room temperature $P_{\rm Na}^{\rm sat}=2\times 10^{-11}$~mbar, with a coefficient that depends on the details of the coverage of the surfaces under vacuum with sodium atoms.  
In presence of LIAD, we can infer the value of the pressure from the loading time of the MOT $\tau_{\rm MOT}$ by using the relation $\tau_{\rm MOT}^{-1}\simeq n\sigma v$, where $n$ is the sodium density in the background vapor, $\sigma$ is the collision cross--section between a trapped atom and an atom from the vapor, and $v\simeq (k_{\rm B} T /m)^{1/2}$ is the average thermal velocity (with $T$ the room temperature and $m$ the mass of an atom).
We assume a typical value $\sigma=10^{-12}$~cm$^{-2}$\cite{cable1990a}, neglecting its energy dependence. The value $\tau_{\rm MOT}=5$~s obtained in presence of LIAD at full power corresponds to a pressure $P_{\rm Na}^{\rm on}\simeq 10\,P_{\rm Na}^{\rm sat}$.
Given the crude assumptions behind our estimate and the fact that the velocity of atoms desorbed by LIAD is not fully thermal~\cite{Gozzini1993a}, this result is in reasonable  
agreement with $\eta = 40$. A final consistency check amounts to calculating the loading rate expected in our MOT from a vapor in equilibrium at room temperature ($T=295$~K). We use the result $R=0.5(P_{\rm Na}^{\rm on}/k_{\rm B}T)V^{2/3}v_{\rm cap}^4(m/2k_{\rm B}T)^{3/2}$ from \cite{monroe1990a}, where $V\simeq 1$~cm$^3$ is the MOT volume and $v_{\rm cap}$ the capture velocity. From a one--dimensional semi--classical analysis for our MOT parameters we deduce $v_{\rm cap}\simeq 35$~m~s$^{-1}$, which leads finally to $R=4\times10^{6}$~s$^{-1}$, in good agreement with the experimental finding. This suggests that the velocity distribution is almost thermal, {\it i.e.} the vapor released via LIAD quickly equilibrates with the walls of the chamber.
  
\subsection{Lifetime in the dipole trap}
\label{DTLt}

Results from the previous section demonstrate that LIAD is an efficient way to increase the partial pressure of sodium in the vacuum chamber. Moreover, LIAD has the important property that it can be controlled on a short time scale. This is a crucial feature for BEC experiments in which evaporation takes place in the same chamber as the MOT. In this case it is mandatory that the increased pressure during MOT loading be transient, and that the background pressure be recovered fast enough to preserve the lifetime of the atoms in the ODT. We infer this lifetime by plotting the number of atoms $N$ in the trap as a function of time (see Fig.(\ref{fig1}{\bf b})). After a fast initial decay for the first $200~$ms, reflecting free evaporation from the ODT~\cite{barrett2001a}, we find that $N$ decreases over a much longer time scale. A fit of the decay of $N$ for $t>200~$~ms by an exponential function $\exp(-t/\tau_{\rm ODT})$ gives $\tau_{\rm ODT}=10.7\pm 1$~s.

The apparent discrepancy between the lifetime in the ODT ($\sim11$~s) and the lifetime in the MOT ($\sim27$~s) can be better understood by analyzing of the loss mechanisms in the ODT. We have performed numerical simulations of the evaporation in the ODT. The simulations simplify the trap geometry to a truncated harmonic trap, but account for 1--body and 3--body collisions~\cite{luiten1996a}, which are significant due to the high density in the trap ($\sim10^{14}$~at~cm$^{-3}$). The rate for these collisions $L_{\rm 3B}=2\times10^{-30}$cm$^6$~s$^{-1}$ has been measured in~\cite{gorlitz2003a} for the $F=1$, $m_{\rm F}=-1$, state and we use this value for all Zeeman states. We derive from these simulations a value for the one--body lifetime due to background collisions $\tau_{\rm 1B}=17\pm 3$~s. The corresponding result for the atom decay is shown by the solid red line in Fig.(\ref{fig1}{\bf b}). This shows that 3--body collisions play a significant role in the ODT, so that the effective decay time $\tau_{\rm ODT}$ is systematically shorter than the one--body decay time $\tau_{\rm 1B}$, and essentially limited by these collisions. We use for simplicity the fitted $\tau_{\rm ODT}$ to analyze the data in the following. The dependence of the lifetime on the background pressure is not qualitatively changed, while this parameter is easier to fit and is model--independent. We attribute the residual discrepancy between $\tau_{\rm 1B}$ and $\tau_{\rm MOT}$ to the largely different depths of the ODT and the MOT, which makes collisions more likely to eject atoms from the former than from the latter.

\begin{figure}[t]
\centering{\includegraphics{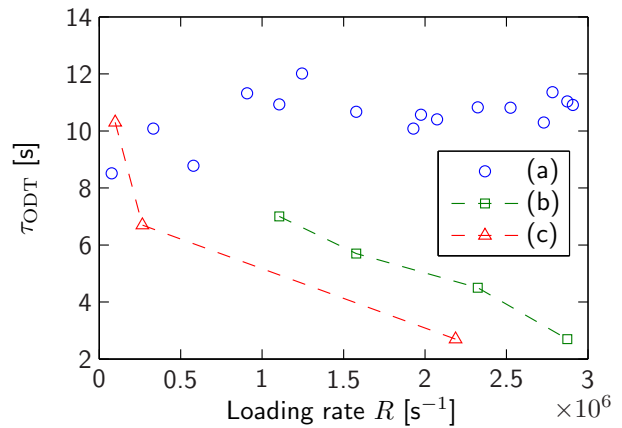}}
\caption{(Color online) MOT loading rate $R$ plotted against $1/\tau_{\rm ODT}$, where $\tau_{\rm ODT}$ is the lifetime in the dipole trap, for experimental situations ({\bf a}): LIAD on for MOT loading, off for dipole trapping; ({\bf b}): LIAD on for both; ({\bf c}): no LIAD, dispensers on for both. Case ({\bf a}) corresponds to a dipole trap lifetime independent of the MOT loading rate, whereas they are inversely related in cases ({\bf b}) and ({\bf c}).}
\label{fig4} 
\end{figure}

In order to determine the effect of LIAD loading on the ODT, we perform lifetime measurements in the ODT for three different cases. In the first case (a), our standard sequence, the LEDs are turned on during the loading of the MOT, and then turned off when the atoms are loaded in the ODT. In the second case (b), the MOT is loaded just as in the first sequence, but the LEDs remain on while the atoms are kept in the ODT, keeping the pressure at the same level as in the MOT loading phase. In the third case (c), LIAD is not used and the pressure of sodium is increased using the sodium dispensers continuously, during both MOT and ODT phases. The different cases are compared by looking at $\tau_{\rm ODT}$ plotted against $R$ in Fig.(\ref{fig4}). In case (a), $\tau_{\rm ODT}$ is independent of $R$, implying that one can load the MOT at high loading rates without deteriorating the lifetime in the ODT. In cases (b) and (c), the behavior is qualitatively different: $\tau_{\rm ODT}$ decreases when the pressure increases and a higher loading rate corresponds to a lower lifetime in the ODT. One can also notice that for a given MOT loading rate, $\tau_{\rm ODT}$ is significantly larger using LIAD than using the dispensers. A probable explanation is that the dispensers are releasing other compounds than sodium  in the UHV chamber when heated, while LIAD is more selective and does not modify the partial pressures of other bodies in a significant way. 

\subsection{Time evolution of pressure}\label{sec:time}
\begin{figure}[t]
\centering{\includegraphics{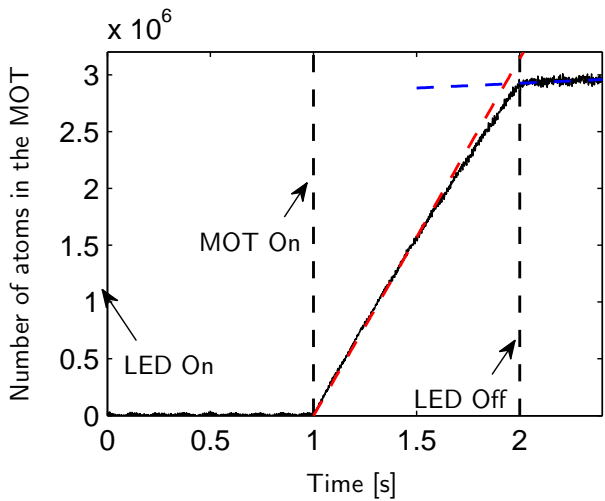}}
\caption{(Color online) Evolution of the loading of the MOT after extinction of the LEDs : the MOT is loaded for $1$~s then the LEDs are turned off. The loading rate shown as a red (blue) dashed line is $3\times 10^6$~s$^{-1}$ ($8\times 10^4$~s$^{-1}$) when the LEDs are on (off).}
\label{fig5}
\end{figure}

The results presented so far show that the sodium partial pressure rapidly drops back to its background level when the LEDs are turned off. This process appears to be fast enough so that the lifetime in the ODT is not diminished. In order to confirm this result and determine the timescale of the decrease in pressure, we perform two measurements on the loading of the MOT. First, the MOT is loaded with LIAD for the first $12$~s, then the LEDs are turned off and the decay of the number of atoms in the MOT is recorded (see Fig.(\ref{fig1}{\bf a})). By fitting the loading and decay phases by two exponentials, we find $1/e$ time constants $\tau_{\rm ON}=6.5\pm 0.5$~s and $\tau_{\rm OFF}=27\pm 1.5$~s with and without LIAD, respectively. Given that $\tau_{\rm OFF}$ is compatible with the loading time of a MOT without LIAD shown in Fig.(\ref{fig3}), it appears that the pressure goes back to the background value on a timescale which is small compared to $\tau_{\rm ON}$. To characterize more accurately how fast this process goes, a second measurement is performed on a shorter timescale, using fluorescence light emitted from the atoms of the MOT to measure the small number of atoms at the very beginning of the loading. We first load the MOT using LIAD for $1~$s, then turn off the LEDs keeping the MOT light on. We observe on Fig.(\ref{fig5}) a sudden change in slope, with a loading rate going from $3\times 10^6$~s$^{-1}$ to $8\times 10^4$~s$^{-1}$, i.e. we recover the same factor $\eta\sim40$ as previously found (see Fig.(\ref{fig3}{\bf a})). This change happens with a characteristic time shorter than $100$~ms, an upper bound limited by the sensitivity of the fluorescence measurement.

\section{Discussion}
\label{sec:disc}

A figure of merit to evaluate the performance of LIAD for preparation of ultracold gases is how low the background pressure in the region of the MOT drops once the desorbing light has been turned off. This depends {\it a priori} not only on the physics of LIAD, but also on technical details such as the effective pumping speed. The different experimental results available in the literature for UHV systems are summarized in table \ref{table2}. The large variations of the reported loading rates can be easily explained with the different parameters of the MOT, in particular with the beam size which impacts the capture velocity. The data about the MOT lifetime depend mainly on the atomic species and on the pressure in the vacuum chamber. The decay of the pressure after turning off the desorbing light reported in this work is among the fastest reported in literature, while the lifetime of the MOT after this extinction is among the longest. Such a slow MOT decay is reported also in~\cite{Du2004,extavour2006a} which are unsurprisingly two other cases in which evaporation takes place in the same spatial region as the one where the MOT is loaded. We find a pressure decay time which is much faster than the other timescales of our experiment, comparable to~\cite{extavour2006a,Zhang2009a}. Our observations are compatible with the scenario where almost all  \na atoms stick to the surfaces of the vacuum system after a few bounces when LIAD is turned off. In the opposite limit where the equilibrium pressure remains large for a long time (longer than $1$~s), the lifetime is limited by the capacity of the pumping system to restore a low background pressure. This may explain the large variations observed in different experiments.

\begin{table*}[htbp]
  \centering
  \begin{tabular}{|c|c|c|c|c|c|c|}
\hline    Atomic & Desorbing & MOT & MOT & Pressure & Conservative & Reference \\
 species & surface & loading rate & lifetime & decay & trap & \\
\hline \hline
\na & TiO$_2$ $+$ SiO$_2$ & $3 \times 10^6$~s$^{-1}$ & $27\,$s & $< 100\,$ms & dipole, $>11$s & this work\\                                               
\hline\hline
$^{87}$Rb & Pyrex & $\sim 10^6$~s$^{-1}$ & $\sim 5$~s & & microchip, $>4\,$s &~\cite{hansel2001a}\\
$^{87}$Rb & SS & $8 \times 10^5$~s$^{-1}$ & & $\sim 100\,$s & none &~\cite{Anderson2001} \\
$^{87}$Rb & PDMS & $2.0 \times 10^8$~s$^{-1}$ & $\sim 10\,$s & & none &~\cite{Atutov2003} \\
$^{87}$Rb & quartz + Pyrex & $\sim 10^6$~s$^{-1}$ & $\sim 30\,$s & $\ll 30\,$s & microchip, $>5\,$s &~\cite{Du2004} \\
$^{87}$Rb & Vycor & $1.2 \times 10^9$~s$^{-1}$ & $\sim 3\,$s & $\sim2\,$s & magnetic &~\cite{Klempt2006} \\
$^{40}$K & Vycor & $8 \times 10^7$~s$^{-1}$ & $\sim 1\,$s & & magnetic &~\cite{Klempt2006} \\
$^{87}$Rb & Pyrex & $3 \times 10^8$~s$^{-1}$ & $\sim 24\,$s& $\sim 100\,$ms& microchip, $>9\,$s&~\cite{extavour2006a}\\
$^{40}$K & Pyrex & $\sim 10^5$~s$^{-1}$ & & & microchip, $>9\,$s& ~\cite{extavour2006a}\\
\na & Pyrex & $4.5 \times 10^7$~s$^{-1}$ & $10\,$s & & none &~\cite{telles2009a}\\
$^{133}$Cs & Quartz & $4 \times 10^3$~s$^{-1}$ & $9.2 $s & $70\ $ms & none &~\cite{Zhang2009a}\\
\hline
  \end{tabular}
  \caption{Summary of available data on LIAD for the preparation of a MOT (including this work). Reported data are best figures in terms of atomic flux. MOT lifetime is measured after switching off the desorbing light. Pressure decay time is defined as the time required for pressure to drop at one tenth of its value during the loading of the MOT. The column labeled ``Conservative trap'' indicates the type of confinement used after the MOT phase. When the trap is realized in the same location as the MOT, we indicate the decay time constant due to collisions with the background gas. SS is Stainless Steel, PDMS is polydimethylsiloxane.}
  \label{table2}
\end{table*}

\section{Conclusion}\label{sec:concl}

In conclusion, we have demonstrated an efficient route to Bose--Einstein condensation of \na in a compact single--chamber setup, with no source of magnetic fields except for the transitory gradient used for the MOT. The MOT is loaded by LIAD, with a steady--state number of atoms of about $2\times 10^{7}$. Its lifetime remains in the order of $30$~s, enough to produce a BEC, thanks to the rapid decrease of the partial pressure of sodium when the desorbing light is switched off. 

The experimental setup described in this work is aimed at rapidly producing small BECs in a single--chamber vacuum system. 
However, relaxing some of the technical constraints imposed on the design of our apparatus may allow to use this technique for producing larger BECs and degenerate Fermi gases via sympathetic cooling. Since we use only 20\% of the available power of our solid--state laser, one could double the size the MOT beams keeping the intensity constant and expect at least an
improvement of a factor 4 in the number of atoms in the MOT~\footnote{In the present apparatus, the beam size is limited by the aperture of the viewports.}. Another very significant gain can be obtained by upgrading the ODT. Since the density in our ODT is limited by inelastic collisions, increasing the trap frequency would probably not be the best option. Enlarging the size of the ODT beams, while keeping the confinement constant would probably be a better route to larger BECs by allowing to catch a higher fraction of atoms from the MOT. This can be done by using more powerful fiber lasers than the one we use, which are commercially available. With these technical improvements, the experimental techniques reported in this work could allow to produce degenerate Fermi gases with atom numbers comparable to what has already been obtained on atom chips~\cite{aubin2005a}.

\begin{acknowledgments}

We wish to thank the members of the ``Bose--Einstein condensates'' and ``Fermi gases'' groups at LKB, as well as J. Reichel and J.H. Thywissen for helpful discussions, and W.L. Kruithof for experimental assistance. LdS acknowledges financial support from the 7th framework programme of the EU, grant agreement number 236240. DJ acknowledges financial support by DGA, contract 2008/450. This work was supported by ANR (``Gascor'' project), IFRAF, by the European Union (MIDAS STREP project), and DARPA (OLE project). Laboratoire Kastler Brossel is a {\it Unit{\'e} mixte de recherche} (UMR n$^{\circ}$ 8552) of CNRS, ENS and UPMC.

\end{acknowledgments}

%

%
%
\end{document}